\documentclass[floats,epsf]{iopart}

\usepackage{iopams}

\begin{document}

\jl{3}

\title[Theory of interlayer exchange interactions]{Theory of interlayer 
exchange interactions in magnetic multilayers}

\author{P Bruno~\footnote[1]{Electronic address: {\tt bruno@mpi-halle.de}}}

\address{Max-Planck-Institut f\"ur Mikrostrukturphysik\\
Weinberg 2, D-06120 Halle, Germany} 


\begin{abstract}
This paper presents a review of the phenomenon of interlayer exchange 
coupling in magnetic multilayers. The emphasis is put on a pedagogical 
presentation of the mechanism of the phenomenon, which has been successfully 
explained in terms of a spin-dependent quantum confinement effect. The 
theoretical predictions are discussed in connection with corresponding 
experimental investigations. \\
$  $\\
published in: J. Phys.: Condens. Matter {\bf 11}, 9403--9419 (1999)
\end{abstract}


\section{Introduction}

Since the first observation by Gr\"unberg {\em et al.\/} \cite{Grunberg1986} 
of antiferromagnetic coupling of Fe films separated by a Cr spacer, the 
interlayer exchange interaction between ferromagnetic layers separated by a 
non-magnetic spacer has been a subject of intense research in the last few 
years. The decisive stimulus came from the discovery, by Parkin 
{\em et al.\/} \cite{Parkin1990}, of {\em oscillations\/} of the interlayer 
exchange coupling in Fe/Cr/Fe and Co/Ru/Co multilayers, as a function of 
spacer thickness. Furthermore, Parkin \cite{Parkin1991} showed that this 
spectacular phenomenon occurs with almost any metal as a spacer 
material. 

This review will be restricted to the case of 
interlayer coupling across non-magnetic metallic spacer layers. This excludes 
the cases of non-metallic spacer \cite{Toscano1992, Mattson1993, Briner1994, 
Inomata1995, Bruno1994, Vries1997}, 
of antiferromagnetic spacer such as Cr or Mn \cite{Ruhrig191, Demokritov1991, 
Unguris1991, Unguris1992, Wolf1993, Fullerton1993, Fullerton1995, 
Meerschaut1995, Schreyer1995, Fullerton1996, Grimditch1996, Schreyer1997}
and of rare-earth 
multilayers \cite{Majkrzak1986, Salamon1986, Majkrzak1991, Rhyne1995}. 
This choice is motivated by the fact that the physical mechanism of the 
coupling in these cases is quite different from the one to be discussed 
here.

The magnetic coupling energy per unit area can usually be expressed 
as
\begin{equation}
E(\theta ) = J \cos\theta ,
\end{equation}
where $\theta$ is the angle between the magnetizations of the two magnetic 
layers, and $J$ is called the interlayer coupling coupling constant. Higher 
order terms in an expansion in powers of $\cos\theta$ are also observed; such 
terms, which give rise to non-collinear alignment of the magnetizations, are 
believed to be of non intrinsic origin and to be related to defects such as 
roughness \cite{Slonczewski1995, Demokritov1998}. Such effects will not be 
considered here. In addition, due to space limitation, the important question 
of the role of alloy disorder and interdiffusion \cite{Bruno1996b, 
Kudrnovsky1996b, Bruno1997, Drchal1998} will not be addressed.

The purpose of this paper is to present as simply as possible the mechanism 
of interlayer exchange coupling in terms of quantum interferences due to 
electron confinement in the spacer layer. The understanding of this mechanism 
relies on ideas due to various authors \cite{Edwards1991, Bruno1991, 
Bruno1993, Stiles1993, Bruno1995}. The presentation give here is based 
on the one of Ref. \cite{Bruno1995}. The emphasis will be on 
physical concepts and pedagogical clarity rather than on mathematical 
rigor. It is organized as 
follows: in the 
next Section, an elementary discussion of quantum confinement is given; in 
Section \ref{sec:IXC}, it is then shown how spin-dependent confinement in the 
spacer layer gives rise to interlayer exchange coupling; Section 
\ref{sec:asympt} is devoted to the limit of large spacer thicknesses, for 
which particularly simple results are obained; 
Sections 
\ref{sec:maglayer} and \ref{sec:overlayer} treat the variation of interlayer 
exchange coupling with magnetic layer thickness and non-magnetic overlayer 
thickness, respectively; finally, 
in Section \ref{sec:strength} 
the strength of the interlayer exchange coupling is discussed in comparison 
with experimental data.

The point of view adopted here reflects the author's subjective views on the 
topic. In particular, due to space limitation, the important literature 
devoted to {\em ab initio\/} 
calculation will not be discussed in detail here. The interested 
reader can find 
complementary information in the various review papers on this subject which 
have been published recently 
\cite{Slonczewski1995, Fert1993, Yafet1994, Fert1995, Vries1996, 
Jones1998, Stiles1999}.

\section{Elementary discussion of quantum confinement} 

For the sake of clarity, we shall first consider an extremely simplified 
model, namely the one-dimensional quantum well, which nevertheless contains 
the essential physics involved in the problem. Then, we shall progressively 
refine the model in order to make it more realistic.

The model consists in a one-dimensional quantum well representing the spacer 
layer (of potential $V=0$ and width $D$), sandwiched between two ``barriers'' 
$A$ and $B$ of respective widths $L_A$ and $L_B$, and respective potentials 
$V_A$ and $V_B$. Note that we use the term ``barrier'' in a general sense, 
i.e., $V_A$ and $V_B$ are not necessarily positive. Furthermore, the barrier 
widths, $L_A$ and $L_B$, can be finite or infinite, without any restriction.

\subsection{Change of the density of states due to quantum interferences}

Let us consider an electron of wavevector $k^+$ (with $k^+ > 0$) propagating 
towards 
the right in the spacer layer; as this electrons arrives on barrier $B$, it 
is partially reflected to the left, with a (complex) anplitude $r_B \equiv 
|r_B| \rme^{\rmi \phi_B}$. The reflected wave of wavevector $k^-$ is in turn 
reflected on barrier $A$ with an amplitude $r_A \equiv |r_A| \rme^{\rmi 
\phi_A}$, an so on.%
\footnote{\ Of course, for the one-dimensional model, one has $k^- = -k^+$; 
however, this property will generally not hold for three-dimensional 
systems to be studied below.}%
\ The module $|r_{A(B)}|$ of the reflection coefficient 
expresses the magnitude of the reflected wave, whereas the argument 
$\phi_{A(B)}$ represents the phase shift due to the reflection (note that 
the latter is not absolutely determined and depends on the choice of the 
coodinate origin).

\begin{figure}
\includegraphics{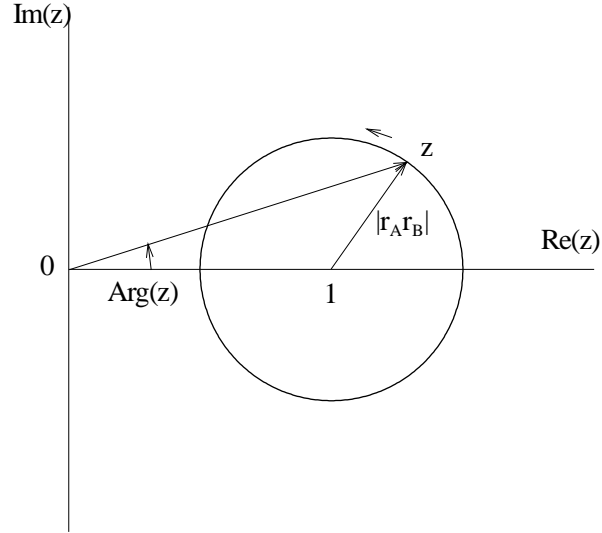}
\vspace*{9cm}
\caption{Graphical interpretation of equation (\protect\ref{eq:th:DeltaN}).}
\label{fig:graphic}
\end{figure}

The interferences between the waves due to the multiple reflections on the 
barriers induce a modification of the density of states in the spacer layer, 
for the electronic state under consideration. The phase shift resulting from 
a complete round trip in the spacer is
\begin{equation}
\Delta \phi = q D + \phi_A + \phi_B \ ,
\end{equation}
with 
\begin{equation}
q \equiv k^+-k^- \ .
\end{equation}
If the interferences are constructive, i.e., if
\begin{equation}
\Delta \phi = 2n\pi
\end{equation}
with $n$ an integer, one has an increase of the density of states; conversely, 
if the interferences are destructive, i.e., if
\begin{equation}
\Delta\phi = (2n+1)\pi 
\end{equation}
one has a reduction of the density of states. Thus, in a first approximation, 
we expect the modification of the density of states in the spacer, 
$\Delta n(\varepsilon )$, to vary with $D$ like
\begin{equation}
\Delta n(\varepsilon) \approx \cos \left( q D + \phi_A +\phi_B \right) .
\end{equation}
Furthermore, we expect that this effect will proportional to the amplitude 
of the reflections on barriers $A$ and $B$, i.e., to $|r_A r_B|$; finally, 
$\Delta n(\varepsilon )$ must be proportional to the width $D$ of the spacer 
and to the density of states per unit energy and unit width,
\begin{equation}
\frac{2}{\pi} \frac{\rmd q }{\rmd\varepsilon}
\end{equation}
which includes a factor of $2$ for spin degenaracy. We can also include the 
effect of higher order interferences due to $n$ round trips in the spacer; the 
phase shift $\Delta \phi$ is then multiplied by $n$ and $|r_A r_B|$ is 
replaced by $|r_A r_B|^n$. Gathering all the terms, we get,
\begin{eqnarray}
\Delta n(\varepsilon ) &\approx& \frac{2D}{\pi}\, 
\frac{\rmd q}{\rmd\varepsilon} \sum_{n=1}^\infty \, |r_Ar_B|^n \, 
\cos n\left( qD +\phi_A +\phi_B \right) \nonumber \\
&=& \frac{2}{\pi} \mbox{ Im} \left( \rmi D\, 
\frac{\rmd q}{\rmd\varepsilon}\, \sum_{n=1}^\infty \, 
\left(r_Ar_B\right)^n \, \rme^{n \rmi qD} \right) \nonumber \\
&=& \frac{2}{\pi} \mbox{ Im} \left( \rmi \, 
\frac{\rmd q}{\rmd\varepsilon}\, \frac{r_Ar_B\, 
\rme^{\rmi qD}}{1 - r_Ar_B\, 
\rme^{\rmi qD}} \right)
\end{eqnarray}
As will appear clearly below, it is more convenient to consider the integrated 
density of states
\begin{equation}
N(\varepsilon ) \equiv \int_{-\infty}^{\varepsilon} n(\varepsilon^\prime )\, 
\rmd\varepsilon^\prime  .
\end{equation}
The modification $\Delta N (\varepsilon)$ of the integrated density of states 
due to electron confinement is
\begin{eqnarray}\label{eq:th:DeltaN}
\Delta N(\varepsilon ) &=& \frac{2}{\pi} \mbox{ Im}\, \sum_{n=1}^\infty \, 
\frac{\left( r_A r_B \right)^n}{n}\, \rme^{n\rmi qD} \nonumber \\
&=& -\, \frac{2}{\pi} \mbox{ Im } \ln \left( 1 -r_A r_B\, \rme^{\rmi qD} 
\right)
\end{eqnarray}
A simple graphical interpretation of the above expression can be obtained by 
noting that $\mbox{Im }\ln (z) = \mbox{Arg }(z)$, for $z$ complex; thus, 
$\Delta N(\varepsilon )$ is given by the argument, in the complex plane, of a 
point located at an angle $\Delta \phi = qD + \phi_A + \phi_B$ on a 
circle of 
radius $|r_A r_B|$ centred in 1. This graphical construction is shown in 
\ref{fig:graphic}.

\begin{figure}
\includegraphics{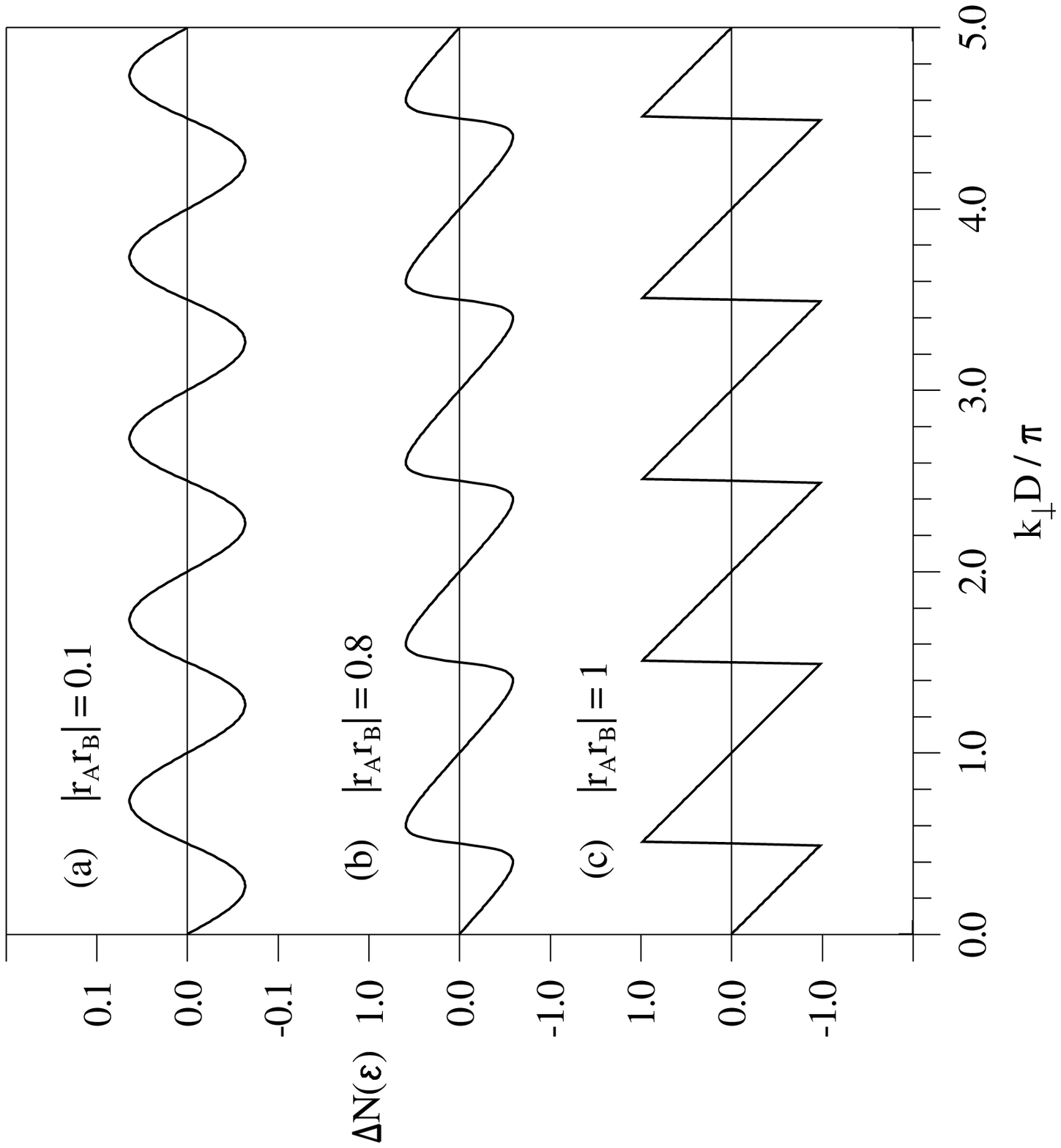}
\vspace*{10cm}
\caption{Variation of $\Delta N(\varepsilon )$ as a function of $D$, for 
various values of the confinement strength: (a) $|r_A r_B| = 0.1$, 
(b) $|r_A r_B| = 0.8$, (c) $|r_A r_B| = 1$ (full confinement). Note the 
different scales along the ordinate axis.}
\label{fig:deltaN}
\end{figure}

The variation of $\Delta N(\varepsilon )$ as a function of $D$ is shown in 
\ref{fig:deltaN}, for various values of the confinement strength $|r_A r_B|$. 
For weak confinement (a), $\Delta N(\varepsilon )$ varies with $D$ in 
sinusoidal manner. As one increases the confinement strength (b), the 
oscillations are distorded, due to higher order interferences. Finally, for 
full confinement (c), $\Delta N(\varepsilon )$ exhibits some jumps that 
correspond to the appearance of bound states. We note however, 
that the period $\Lambda$ of 
the oscillations of $\Delta N(\varepsilon)$ does not depend on the confinement 
strength, but only on the wavevector $q \equiv k^+-k^-$, namely, 
$\Lambda = 2\pi /q$.

So far, we have implicitely restricted ourselves to positive energy states. 
Negative energy states (i.e., of imaginary wavevector) are forbidden in 
absence of the barriers $A$ and $B$, because their amplitude diverges either 
on the right hand side or on the left hand side, so that they cannot be 
normalized. This matter of fact no longer holds in the presence of the 
barriers if $V_A$ (or $V_B$, or both $V_A$ and $V_b$) is negative: 
the negative energy states, 
i.e., varying exponentially in the spacer, can be connected to allowed states 
of $A$ or $B$. In order to treat these states consistently, we simply have to 
extend the concept of reflection coefficient to to states of imaginary 
wavevector, which is straightforward. One can check that, with this 
generalization, (\ref{eq:th:DeltaN}) acounts properly for the contribution of 
the evanescent states. Physically, this can be interpretated as a coupling of 
$A$ and $B$ by tunnel effect \cite{Bruno1994, Bruno1995}.. 

\subsection{Energy associated with the quantum interferences in the 
spacer}

Let us now study the modification of the energy of the system due to the 
quantum interferences. In order to conserve the total number of electrons, 
it is convenient to work within the grand-canonical ensemble, and to consider 
the thermodynamic grand-potential, which is given by
\begin{eqnarray}
\Phi &\equiv& - k_BT \int_{-\infty}^{+\infty} \ln \left[ 1+ \exp 
\left( \frac{\varepsilon_F - \varepsilon}{k_BT}\right) \right] \, 
n(\varepsilon )\, \rmd\varepsilon \nonumber \\
&=& - \int_{-\infty}^{+\infty} N(\varepsilon )\, f(\varepsilon)\, 
\rmd\varepsilon .
\end{eqnarray}
At $T=0$, this reduces to 
\begin{eqnarray}
\Phi &\equiv& \int_{-\infty}^{\varepsilon_F} \left( \varepsilon - 
\varepsilon_F \right)\, n(\varepsilon )\, \rmd\varepsilon \nonumber \\
&=& -\int_{-\infty}^{\varepsilon_F} N(\varepsilon)\, \rmd\varepsilon .  
\end{eqnarray}
The energy $\Delta E$ associated with the interferences is the contribution 
to $\Phi$ corresponding to $\Delta N(\varepsilon )$, 
\begin{equation}
\Delta E = \frac{2}{\pi} \mbox{ Im} \int_{-\infty}^{+\infty} \ln 
\left( 1 - r_Ar_B \, \rme^{\rmi qD} \right) \, \rmd\varepsilon .
\end{equation}

\subsection{Three-dimensional layered system}

The generalization of the above discussion to the more realistic case of a 
three-dimensional layered system is immediate. Since the system is 
invariant by translation parallely to the plane, the in-plane 
wavevector $\bf{k}_\|$ is a good quantum number. Thus, for a given 
$\bf{k}_\|$, one has an effective one-dimensional problem analogous to the 
one discussed above. The resulting effect of quantum intereferences is 
obtained by summing on $\bf{k}_\|$ over the two-dimensional Brillouin zone. 
The modification of integrated density 
of states per unit area is
\begin{equation}
\Delta N(\varepsilon ) = -\,\frac{1}{2\pi^3}\mbox{ Im} \int
\rmd^2\mbox{\bf k}_\|  \, \ln\left( 1- r_Ar_B\, \rme^{\rmi q_\bot D} \right) \ ,
\end{equation}
and the interference energy per unit area is 
\begin{equation}\label{eq:th:QSE-Energy-3D}
\Delta E = \frac{1}{2\pi^3}\mbox{ Im} \int \rmd^2\mbox{\bf k}_\|
\int_{-\infty}^{+\infty}f(\varepsilon )\, \ln\left( 1- r_Ar_B\, 
\rme^{\rmi q_\bot D}
\right) \, \rmd\varepsilon \ .
\end{equation}

\subsection{Quantum size effect in an overlayer}

The case of a thin overlayer deposited on a substrate is of considerable 
interest. In this case, one of the barriers (say, $A$) consists of the vacuum, 
and barrier $B$ is constituted by the substrate itself. The potential of the 
vacuum barrier is $V_{\rm vac} = \varepsilon_F + W$, where $W$ is the the work 
function; thus it is perfectly reflecting for occupied states, i.e., 
$|r_{\rm vac}| = 1$. On the other hand, the reflection on the substrate (or 
coefficent $r_{\rm sub}$) may be total or partial, depending on the band 
matching for the state under consideration.

The spectral density of the occupied states in the overlayer can be 
investigated experimentally by photoemission spectroscopy; in addition, by 
using inverse photoemission, one can study the unoccupied states. If 
furthermore these techniques are used in the ``angle-resolved'' mode, they 
give information on the spectral density \emph{locally in the $\bf{k}_\|$ 
plane}.

For a given thickness of the overlayer, the photoemission spectra (either 
direct or inverse) exhibit some maxima and minima corresponding, respectively, 
to the energies for which the interferences are constructive and destructive. 
When the confinement is total, narrow peaks can be observed, which correspond 
to the quantized confined states in the overlayer, as was pointed out by Loly 
and Pendry \cite{Loly1983}.

Quantum size effects due to electron confinement in the photoemission spectra 
of overlayers have been observed in various non-magnetic systems 
\cite{Wachs1986, Lindgren1987, Lindgren1988, Lindgren1988b, Lindgren1989, 
Miller1988, Mueller1989, Mueller1990, Jalochowski1992}. In particular, the 
systems Au(111)/Ag/vacuum and Cu(111)/Ag/vacuum offer excellent examples of 
this phenomenon \cite{Miller1988, Mueller1990}.

\subsection{Paramagnetic overlayer on a ferromagnetic substrate: 
Spin-polarized quantum size effect} 

So far our discussion concerned exclusively non-magnetic systems. 
Qualitatively new behavior can be expected when some of the layers are 
ferromagnetic. A case of particular interest is the one of a paramagnetic 
overlayer on a ferromagnetic substrate. 

In the interior of the overlayer, the potential is independent of the spin; 
therefore the propagation of electrons is described by a wavevector $k_\bot$ 
which is spin-independent. The reflection coefficient on the vacuum barrier, 
$r_{\rm vac}$, is also spin-independent. However, the ferromagnetic substrate 
constitutes a spin-dependent potential barrier; thus, the substrate reflection 
coefficients for electrons with a spin parallel to the majority and minority 
spin directions of the substrate, respectively $r^\uparrow_{\rm sub}$ and 
$r^\downarrow_{\rm sub}$. It is convenient to define the spin average 
\begin{equation}
\overline{r}_{\rm sub} \equiv \frac{r_{\rm sub}^\uparrow + 
r_{\rm sub}^\downarrow}{2}
\end{equation}
and the spin asymmetry 
\begin{equation}
\Delta r_{\rm sub} \equiv \frac{r_{\rm sub}^\uparrow - 
r_{\rm sub}^\downarrow}{2} .
\end{equation}
In this case, the electron confinement in the overlayer gives rise to a 
spin-dependent modulation of the spectral density versus overlayer thickness; 
the period of the modulation is the same for both spins, whereas the amplitude 
and phase are expected to be spin-dependent. 

The quantum size effects in paramagnetic overlayers on a ferromagnetic 
substrate have been investigated by several groups \cite{Brookes1991, 
Ortega1992, Ortega1993, Ortega1993b, Garrison1993, Carbone1993, Smith1994, 
Johnson1994, Himpsel1995, Crampin1996, Segovia1996, Klaesges1998, 
Kawakami1998, Kawakami1999}. The 
systems studied most are Cu overlayers on a Co(001) substrate and Ag 
overlayers on a Fe(001) substrate. 
Ortega and Himpsel \cite{Ortega1992, Ortega1993} observed a quantum size 
effect in the normal-emission photoelectron spectra of copper overlayer on 
fcc cobalt (001) substrate. They observed peaks due to quantum size effects 
both in the photoemission and in the inverse photoemission spectra an 
oscillation of the photoemission intensity. These quantum size effects 
manifest themselves also by an oscillatory behavior of the photoemission 
intensity at the Fermi level; as the observed oscillation period (5.9~atomic 
layers) is close to the long period of interlayer exchange coupling 
oscillations in Co/Cu(001)/Co, they suggested that the two phenomena should be 
related to each other; they also claimed that the observed oscillations in 
photoemission are spin dependent and due mostly to minority electrons. A 
direct confirmation of this conjecture has been given independently by 
Garrison {\em et al.\/} \cite{Garrison1993} and by Carbone {\em et al.\/} 
\cite{Carbone1993} 
by means of spin-polarized photoemission. They found that both the 
intensity and the spin-polarization exhibit an oscillatory behavior with the 
same period (5 -- 6 atomic layers), but have opposit phases, which 
indicates that the quantum size effect does indeed take place predominantly 
in the minority-spin band as proposed by Ortega and Himpsel \cite{Ortega1992, 
Ortega1993}. 
Recently, Kl\"asges {\em et al.\/} \cite{Klaesges1998} and 
Kawakami {\em et al.\/} \cite{Kawakami1999} have observed 
spin-polarized 
quantum size effects in a copper overlayer on cobalt (001) for a non-zero 
in-plane wavevector corresponding to the short period oscillation of 
interlayer exchange coupling in Co/Cu(001)/Co; they observed short-period 
oscillations of the photoemission intensity in good agrement with the 
short-period oscillations of interlayer coupling. This observation provides 
a further confirmation of the relation between quantum size effects in 
photoemission and oscillation of interlayer exchange coupling. 

Photoemission studies of quantum size effects have also been performed in 
other kinds of systems such as ferromagnetic overlayer on a non-magnetic 
substrate, or systems comprising more layers \cite{Himpsel1991, Ortega1993c, 
Li1995, Himpsel1995b, Li1997}. 

Photoemission spectroscopy undoubtly constitutes a method of choice for 
investigating quantum size effects in metallic overlayers: this is due to its 
unique features, which allow selectivity in energy, in-plane wavevector, and 
spin. 

Besides photemission, spin-polarized quantum size effects in paramagnetic 
overlayers on a ferromagnetic substrate are also responsible for oscillatory 
behavior (versus overlayer thickness) of spin-polarized secondary electron 
emission \cite{Koike1994, Furukawa1996}, linear \cite{Bennett1990, 
Katayama1993, Carl1995, Megy1995, Bruno1996c, Suzuki1998} and non-linear 
\cite{Luce1996, Kirilyuk1996} magneto-optical Kerr effect, 
and magnetic anisotropy 
\cite{Weber1996, Back1997}. However, these effects usually involve a 
summation over all electronic states, so that the quantitative analysis of 
the quantum size effects may be fairly complicated.

\section{Interlayer exchange coupling due to quantum interferences}
\label{sec:IXC}

Let us now consider the case of a paramagnetic layer sandwiched between two 
ferromagnetic barriers $A$ and $B$. Now, the reflection coefficients on both 
sides of the paramagnetic spacer layer are spin dependent. \emph{A priori} the 
angle $\theta$ between the magnetizations of the two ferromagnetic can take 
any value; however, for the sake of simplicity, we shall restrict ourselves 
here to the ferromagnetic (F) configuration (ie., $\theta =0$) and the 
antiferromagnetic (AF) one (i.e., $\theta = \pi$).  

For the ferromagnetic configuration, the energy change per unit area due 
to quantum 
interference is easily obtained from (\ref{eq:th:QSE-Energy-3D}), i.e.,
\begin{eqnarray}
\Delta E_F &=& \frac{1}{4\pi^3}\mbox{ Im} \int \rmd^2 {\bf k}_\|
\int_{-\infty}^{+\infty} f(\varepsilon ) \nonumber \\
&& \times  \left[ \ln\left( 1- r_A^\uparrow r_B^\uparrow
\rme^{\rmi q_\bot D} \right) 
+ \ln\left( 1- r_A^\downarrow
r_B^\downarrow \rme^{\rmi q_\bot D} \right) \right] \, \rmd\varepsilon \ .
\end{eqnarray}
In this equation, the first and the second term correspond respectively to 
majority- and  minority-spin electrons. The antiferromagnetic conguration is 
obtained by reversing the magnetization of $B$, i.e., by interchanging 
$r_B^\uparrow$ and $r_B^\downarrow$; thus the corresponding energy per unit 
area is
\begin{eqnarray}
\Delta E_{AF} &=& \frac{1}{4\pi^3}\mbox{ Im} \int \rmd^2 {\bf k}_\|
\int_{-\infty}^{+\infty} f(\varepsilon ) \nonumber \\
&& \times  \left[ \ln\left( 1- r_A^\uparrow r_B^\downarrow
\rme^{\rmi q_\bot D} \right) 
+ \ln\left( 1- r_A^\downarrow
r_B^\uparrow \rme^{\rmi q_\bot D} \right) \right] \, \rmd\varepsilon \ .
\end{eqnarray}
Thus, the interlayer exchange coupling energy is
\begin{eqnarray}
E_F - E_{AF} &=& \frac{1}{4\pi^3}\mbox{ Im} \int \rmd^2 {\bf k}_\|
\int_{-\infty}^{+\infty}f(\varepsilon ) \nonumber \\
&& \times \ln\left[ \frac{ \left( 1-
r_A^\uparrow r_B^\uparrow \rme^{\rmi q_\bot D} \right)\left( 1-
r_A^\downarrow r_B^\downarrow \rme^{\rmi q_\bot D} \right)}{ \left( 1-
r_A^\uparrow r_B^\downarrow \rme^{\rmi q_\bot D} \right)\left( 1-
r_A^\downarrow r_B^\uparrow \rme^{\rmi q_\bot D} \right)}\right] \,
\rmd\varepsilon 
\end{eqnarray}
which can be simplified as 
\begin{equation}
E_F - E_{AF} \approx -\ \frac{1}{\pi^3}\mbox{ Im} \int \rmd^2 {\bf k}_\|
\int_{-\infty}^{\infty}f(\varepsilon )\, \Delta r_A \Delta r_B\,
\rme^{\rmi q_\bot D}  \, \rmd\varepsilon 
\end{equation}
in the limit of weak confinement. The above expression for the IEC has a rather 
transparent physical interpretation. First, as the integrations on 
{\bf k}$_\|$ over the first two-dimensional Brillouin zone and on the energy up 
to the Fermi level show, the IEC is a sum of contributions from all occupied 
electronic states. The contribution of a given electronic state, of energy 
$\varepsilon$ and in-plane wavevector {\bf k}$_\|$, consists of the product 
of three factors: the two factors $\Delta r_A$ and $\Delta r_B$ express the 
spin-asymmetry of the confinement due to the magnetic layers $A$ and $B$, 
respectively, while 
the exponential factor $e^{iq_\bot D}$ describes the propagation 
through the spacer and is responsible for the interference (or quantum size) 
effect. Thus, this approach establishes an explicit and direct link between 
oscillatory IEC and quantum size effects such as observed in photoemission.

\section{Asymptotic behavior for large spacer thicknesses}\label{sec:asympt}

In the limit of large spacer thickness $D$, the exponential factor oscillates 
rapidly with $\varepsilon$ and {\bf k}$_\|$, which leads to substantial 
cancellation 
of the contributions to the IEC due to the different electronic states. 
However, because the integration over energy is abruptly stopped at 
$\varepsilon_F$, states located at the Fermi level give predominant 
contributions. Thus the integral on $\varepsilon$ may be calculated
by fixing all other factors to their value at $\varepsilon_F$,
and by expanding $q_\bot \equiv k_\bot^+ -k_\bot^-$ around
$\varepsilon_F$, i.e.,
\begin{equation}
q_\bot \approx q_{\bot F} +2 \frac{\varepsilon -\varepsilon_F}
{\hbar v_{\bot F}^{+-}} ,
\end{equation}
with
\begin{equation}
\frac{2}{v_{\bot F}^{+-}} \equiv \frac{1}{v_{\bot F}^+} -
\frac{1}{v_{\bot F}^-} .
\end{equation}
The integration (see Ref.~\cite{Bruno1995} for details) yields
\begin{eqnarray}
E_F-E_{AF} &=& \frac{1}{2\pi^3} \mbox{ Im} \int d^2\mbox{\bf k}_\|\,
\frac{i\hbar v_{\bot F}^{+-}}{D}\, \Delta r_A\Delta r_B
e^{iq_{\bot F}D} \nonumber \\
&&\times F(2\pi\, k_BT\,D/\hbar v_{\bot F}^{+-}) ,
\end{eqnarray}
where
\begin{equation}
F(x) \equiv \frac{ x}{\sinh x} .
\end{equation}
In the above equations, $q_{\bot F}$ is a vector spanning the
{\em complex Fermi surface\/}; the velocity $v_{\bot F}^{+-}$ is
a combination of the group velocities at the points 
$({\bf k}_\|, k_{\bot F}^+)$ and $({\bf k}_\|, k_{\bot F}^-)$ of the Fermi 
surface.

Next, the integration on $\mbox{\bf k}_\|$ is performed by noting,
that, for large spacer thickness $D$, the only significant
contributions arise from the neighboring of critical vectors
$\mbox{\bf k}_\|^\alpha$ for which $q_{\bot F}$ is stationary. Around such
vectors, $q_{\bot F}$ may be expanded as
\begin{equation}
q_{\bot F} = q_{\bot F}^\alpha - \frac{ \left( k_x-k_x^\alpha
\right)^2} {\kappa_x^\alpha} - \frac{ \left( k_y-k_y^\alpha
\right)^2} {\kappa_y^\alpha} ,
\end{equation}
where the crossed terms have been canceled by a proper choice of
the $x$ and $y$ axes; $\kappa_x^\alpha$ and $\kappa_y^\alpha$
are combinations of the curvature radii of the Fermi
surface at $(\mbox{\bf k}_\|^\alpha ,k_{\bot}^{+\alpha})$ and 
$(\mbox{\bf
k}_\|^\alpha ,k_{\bot}^{-\alpha})$. 

The integral is calculated by using the stationary phase
approximation \cite{Bruno1995}, and one obtains
\begin{eqnarray}\label{eq:J_asympt}
E_F-E_{AF} &=& \mbox{ Im } \sum_\alpha \frac{ \hbar v_\bot^\alpha
\kappa_\alpha}{2\pi^2D^2} \Delta r_A^\alpha \Delta r_B^\alpha
e^{iq_\bot^\alpha D}\nonumber \\
&&\times F(2\pi k_BTD/\hbar v_\bot^\alpha ) ,
\end{eqnarray}
where $q_\bot^\alpha$, $v_\bot^\alpha$, $\Delta r_A^\alpha$,
$\Delta r_B^\alpha$ correspond to the critical vector $\mbox{\bf
k}_\|^\alpha$, and
\begin{equation}
\kappa_\alpha \equiv  \left(\kappa_x^\alpha\right)^{1/2}
\left(\kappa_y^\alpha \right)^{1/2} ;
\end{equation}
in the above equation, one takes the square root with an argument
between $0$ and $\pi$.

\begin{figure}[ht] 
\includegraphics{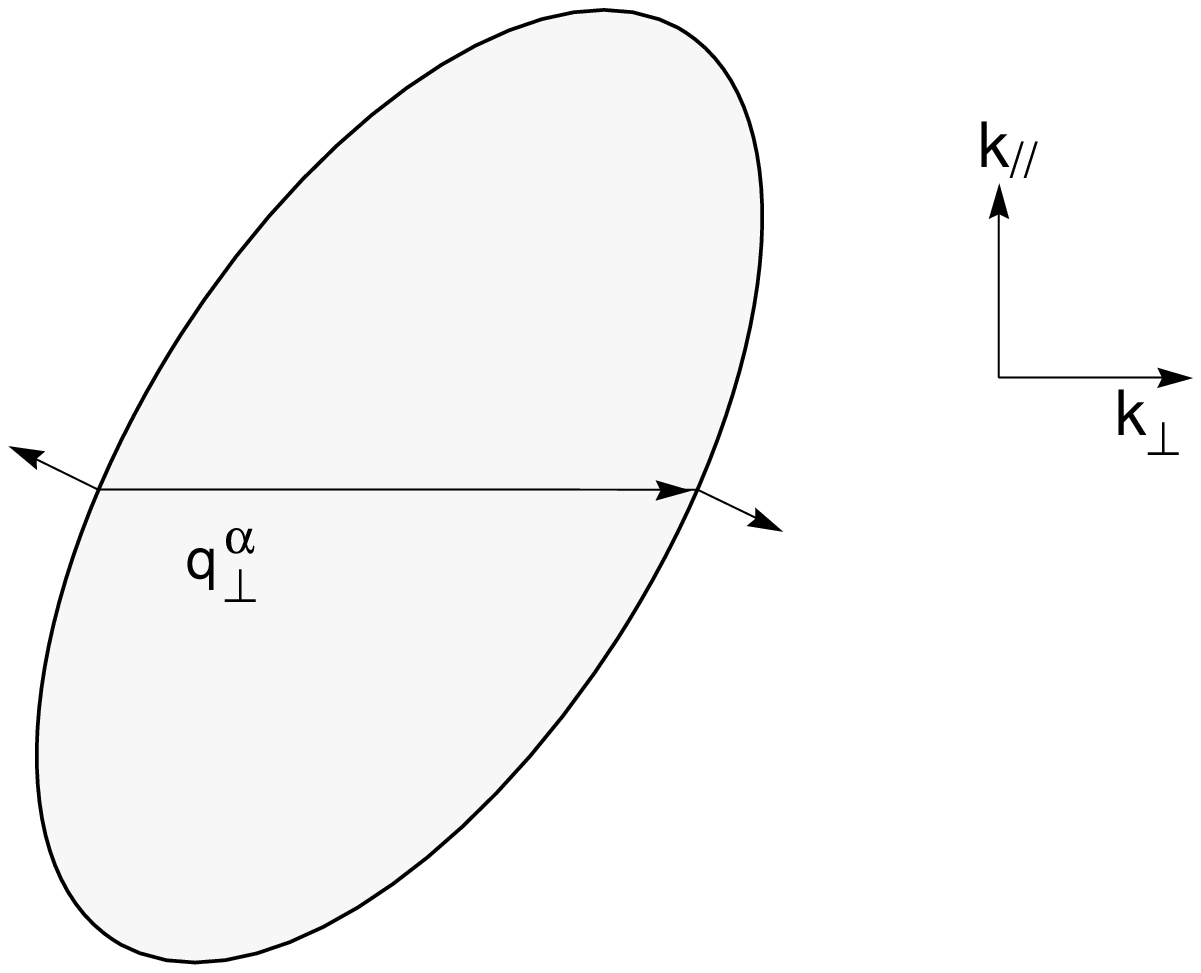}
\vspace*{8cm}
\caption{Sketch showing the wavevector $q_\bot^\alpha$ giving the oscillation
period of oscillatory interlayer exchange coupling, for the case of a 
non-spherical Fermi 
surface.}
\label{ fig_spanning}
\end{figure}

This analysis shows 
that {\em in fine\/}, the only remaining contributions in the limit 
of large spacer 
thickness $D$ arise from the neighborhood of states having in-plane wavevectors 
{\bf k}$_\|^\alpha$ such that the spanning vector of the Fermi surface 
$q_{\bot F} = k_{\bot F}^+ - k_{\bot F}^-$ is stationary with respect to 
{\bf k}$_\|$ for {\bf k}$_\| =${\bf k}$_\|^\alpha$, and the corresponding 
contribution oscillates with a wavevector equal to $q_{\bot F}^\alpha$. This 
selection rule was first derived in the context of the RKKY model 
\cite{Bruno1991}; it is illustrated in Fig.~\ref{ fig_spanning}. 
There may be several such stationary spanning vectors and, hence, 
several oscillatory components; they are labelled by the index $\alpha$.

\begin{figure}[h]
\includegraphics{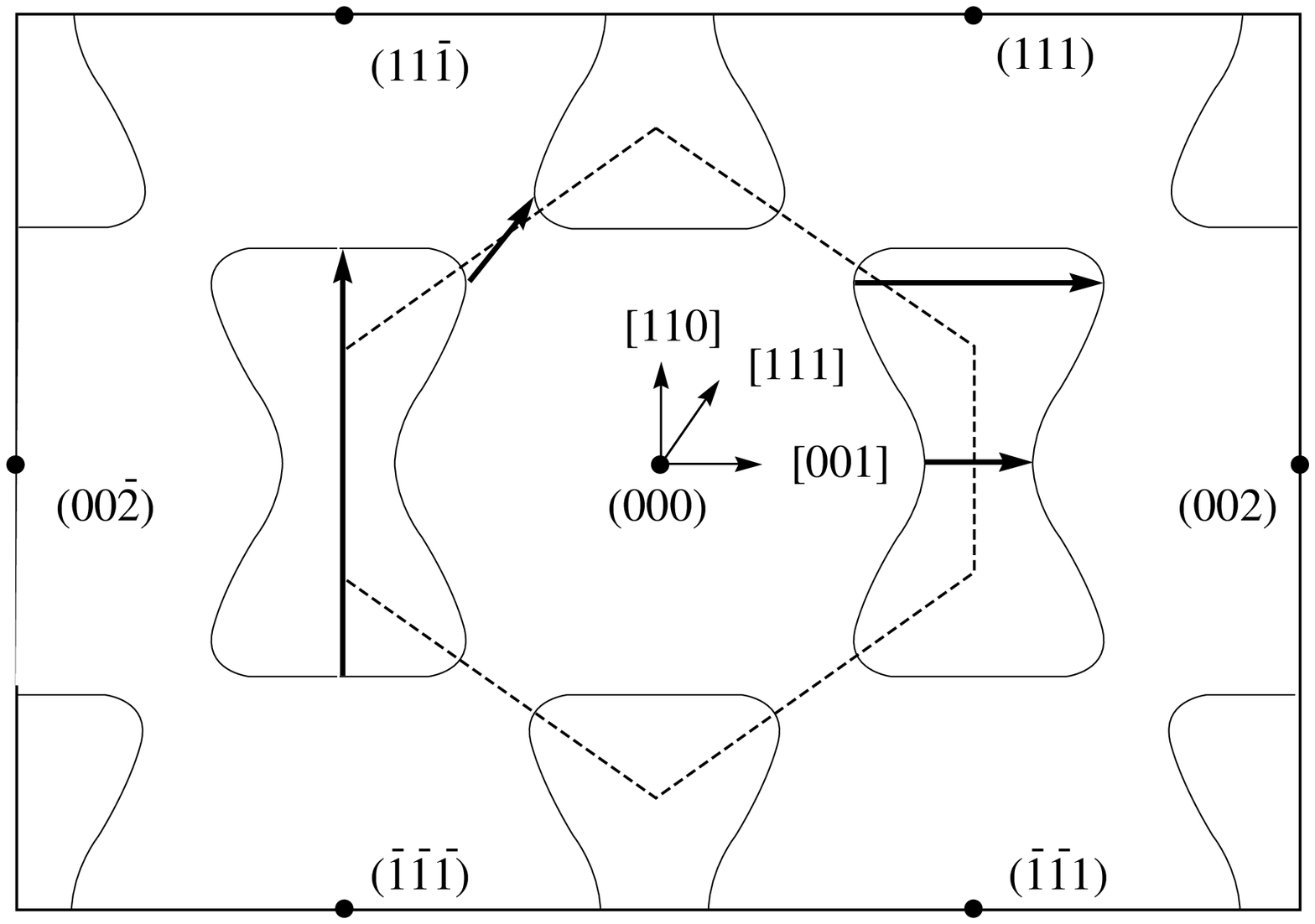}
\vspace*{9cm}
\caption{Cross section of the Fermi surface of Cu along the $(1\bar 10)$ plane 
passing through the origin. The solid dots indicate the reciprocal alttice 
vectors. The dashed lines indicate the boundary of the first brilouin zone. 
The solid arrows, respectiveley horizontal, oblique, and vertical, indicate the
vectors $q_\bot^\alpha$ giving the oscillation period(s), respectively for the 
$(001)$, $(111)$, and $(110)$ orientations.}
\label{ fig_Fermi_Cu}
\end{figure}

\begin{table}
\caption{Comparison between the theoretical predictions of 
Ref.~\protect\cite{Bruno1991} and 
experimental observations for the oscillation periods of interlayer 
exchange coupling versus spacer thickness.}
\label{tab:period}
\vspace*{0.5\baselineskip}
\lineup
\begin{tabular}{@{}c|l|llc@{}}
\br
spacer & theory & system & \0\0experiment & Ref. 
\\
\mr
        &                        & Co/Cu/Co(111) & 
\0\0$\Lambda_{\0} \approx$ 5 AL 
& \cite{Parkin1991} \\
\cline{3-5}
Cu(111) & $\Lambda_{\0} =$ 4.5 AL & Co/Cu/Co(111) & 
\0\0$\Lambda_{\0} \approx$
6 AL & \cite{Mosca1991}  \\
\cline{3-5}
     &                            & Co/Cu/Co(111) & 
\0\0$\Lambda_{\0} \approx$
4.5 AL & \cite{Schreyer1993}  \\
\cline{3-5}
        &                        & Fe/Cu/Fe(111) & 
\0\0$\Lambda_{\0} \approx$
6 AL & \cite{Petroff1991} \\
\mr
        &                        & Co/Cu/Co(001) & 
\0\0$\Lambda_{\0} \approx$ 6 AL 
& \cite{Miguel1991} \\
\cline{3-5}
        &                      & Co/Cu/Co(001) & 
\0\0$\Lambda_1 \approx$ 
2.6 AL & \cite{Johnson1992}  \\
        &                      &               & 
\0\0$\Lambda_2 \approx$ 
8 AL &                       \\
\cline{3-5}
Cu(001) & $\Lambda_1 =$ 2.6 AL & Co/Cu/Co(001) & 
\0\0$\Lambda_1 \approx$ 
2.7 AL & \cite{Weber1995} \\
        & $\Lambda_2 =$ 5.9 AL &               & 
\0\0$\Lambda_2 \approx$
6.1 AL \\
\cline{3-5}
        &                        & Co/Cu/Co(001) & 
\0\0$\Lambda_1 \approx$ 
2.7 AL & \cite{Kawakami1999} \\
        &                        &               & 
\0\0$\Lambda_2 \approx$
5.6 AL \\
\cline{3-5}
        &                        & Fe/Cu/Fe(001) & 
\0\0$\Lambda_{\0} \approx$ 7.5 AL 
& \cite{Bennett1990} \\
\mr
Ag(001) & $\Lambda_1 =$ 2.4 AL & Fe/Ag/Fe(001) & 
\0\0$\Lambda_1 \approx$ 
2.4 AL & \cite{Unguris1993}  \\
        & $\Lambda_2 =$ 5.6 AL &               & 
\0\0$\Lambda_2 \approx$ 
5.6 AL &                       \\
\mr
        &                      & Fe/Au/Fe(001) & 
\0\0$\Lambda_1 \approx$ 
2 AL & \cite{Fuss1992}  \\
Au(001) & $\Lambda_1 =$ 2.5 AL &               & 
\0\0$\Lambda_2 \approx   
7-8$ AL &                       \\
\cline{3-5}
        & $\Lambda_2 =$ 8.6 AL & Fe/Au/Fe(001) & 
\0\0$\Lambda_1 \approx$ 
2.5 AL & \cite{Unguris1994, Unguris1997}  \\
        &                          &               & 
\0\0$\Lambda_2 \approx$ 
8.6 AL &                       \\
\br
\end{tabular}
\end{table}

The above selection rule allows to predict the oscillation period(s) of the 
interlayer exchange coupling versus spacer thickness just by inspecting the 
bulk Fermi surface of the spacer material. In view of an experimental test of 
these predictions, noble metal spacer layers appear to be the best suited 
candidates; there are several reasons for this choice:
\begin{itemize}
\item Fermi surfaces of noble metals are known very accurately from de Haas-van 
Alphen and cyclotron resonance experiments \cite{Halse1969};
\item since only the $sp$ band intersect the Fermi level, the Fermi surface is 
rather simple, and does not depart very much from a free-electron Fermi sphere;
\item samples of very good quality with noble metals as a spacer layer could be 
prepared.
\end{itemize}

Fig.~\ref{ fig_Fermi_Cu} shows a cross-section of the Fermi surface of Cu, 
indicating the stationary spanning vectors for the (001), (111), and (110) 
crystalline orientations \cite{Bruno1991}; the Fermi surfaces of Ag and Au are 
qualitatively similar. For the (111) orientation, a single (long) period is 
predicted; for the (001) orientation, both a long period and a short period are 
predicted; for the (110) orientation, four different periods are predicted 
(only one stationary spanning vector is seen in figure \ref{ fig_Fermi_Cu}, 
the three others being located in other cross-sections of the Fermi surface). 
These theoretical predictions have been confirmed successfully by numerous 
experimental observations. In particular, the coexistence of a long and a short 
period for the (001) orientation has been confirmed for Cu 
\cite{Vries1996, Kawakami1999, 
Johnson1992, Johnson1993, 
Weber1995}, 
Ag \cite{Unguris1993}, and Au \cite{Fuss1992, Unguris1994, Unguris1997}; 
and the experimental periods have been found to be in excellent agreement 
with the theoretical ones. A comparison of the theoretically predicted 
oscillation periods and the experimentally observed ones is given in Table 
\ref{tab:period}.

In a further attempt to test the theoretical predictions for the periods of 
oscillatory coupling, several groups \cite{Okuno1993, Parkin1993, Bobo1993}
have undertaken to modify in a controlled manner the size of the Fermi surface 
(and hence, the period of the coupling) by alloying the spacer noble metal (Cu) 
with a metal of lower valence (Ni); in both cases, the change in oscillation 
period due to alloying has been found in good agreement with the expected 
change in the Fermi surface. 

\section{Effect of magnetic layer thickness}\label{sec:maglayer}

As already mentioned, the influence of the IEC on the ferromagnetic layer 
thickness is contained in the reflection coefficients $\Delta r_A$ and 
$\Delta r_B$. If the ferromagnetic layers are of finite thickness, reflections 
usually may take place at the two interfaces bounding the ferromagnetic layers, 
giving rise to interferences \cite{Bruno1993b}, and hence, to oscillations of 
the IEC versus ferromagnetic layers thickness. A more detailled discussion of 
this effect is given in Refs. \cite{Bruno1995, Bruno1993b}. This behavior was 
first predicted from calculations based upon a free-electron model 
\cite{Barnas1992}. 
The amplitude of the 
oscillations of the IEC versus ferromagnetic layers thickness is generally much 
smaller than the oscillations versus spacer thickness, and do not give rise to 
changes of sign of the IEC. 
On the experimental point of view, this effect 
was confirmed by Bloemen {\em et al.\/} \cite{Bloemen1994} in Co/Cu/Co(001) 
and by Back {\em et al.\/} \cite{Back1995} in Fe/Cu/Co(001). It 
has also been confirmed theoretically by Nordstr\"om {\em et al.\/} 
\cite{Nordstrom1994}, Lang {\em et al.\/} \cite{Lang1996}, Drchal 
{\em et al.\/} \cite{Drchal1996}, and by Lee and Chang \cite{Lee1996}.

\section{Effect of overlayer thickness}\label{sec:overlayer}

A more (at first sight) surprising behavior is the dependence of the IEC on 
the thickness of an external overlayer. From a na\"\i ve point of view, one 
might think that layers external to the basic ferromagnet/spacer/ferromagnet 
sandwich should not influence the interaction between the two ferromagnetic 
layers. This view is incorrect, in particular when the system is covered by an 
ultrathin protective overlayer. In this case, the electrons are able to reach 
the vacuum barrier, which is a perfectly reflecting one, so that strong 
confinement and interference effects take place in the overlayer, which leads 
to a weak but significant oscillatory variation of the IEC as a function of 
the overlayer thickness \cite{Bruno1996}. 

\begin{table}
\caption{Comparison between the theoretical predictions of 
Ref.~\protect\cite{Bruno1996} and 
experimental observations for the oscillation periods of interlayer 
exchange coupling versus overlayer thickness.}
\label{ tab_comp}
\vspace*{0.5\baselineskip}
\lineup
\begin{tabular}{@{}c|l|llc@{}}
\br
overlayer & theory & system & \0\0experiment & Ref. 
\\
\mr
Cu(001) & $\Lambda_1 =$ 2.6 AL & Cu/Co/Cu/Co/Cu(001) & 
\0\0$\Lambda_{\0} 
\approx$ 5 AL & \cite{Vries1995} \\
        & $\Lambda_2 =$ 5.9 AL &                      &      & \\
\mr
Au(001) & $\Lambda_1 =$ 2.5 AL & Au/Fe/Au/Fe/Au(001) & 
\0\0$\Lambda_1 
\approx$ 2.6 AL & \cite{Okuno1995} \\
        & $\Lambda_2 =$ 8.6 AL &                      & 
\0\0$\Lambda_2 
\approx$ 8.0 AL & \\
\mr
Au(111) & $\Lambda_{\0} =$ 4.8 AL & Au/Co/Au/Co/Au(111) & 
\0\0$\Lambda_{\0} 
\approx $ 5 AL & \cite{Bounouh1996} \\
\br
\end{tabular}
\end{table}

This effect, which follows directly from the quantum 
interference (or quantum size effect) mechanism, has been proposed and 
experimentally confirmed independently by de Vries {\em et al.\/} 
\cite{Vries1995} for the Co/Cu/Co(001) system with a Cu(001) overlayer, 
by Okuno and Inomata \cite{Okuno1995} for the Fe/Au/Fe(001) system with a 
Au(001) overlayer, and by Bounouh {\em et al.\/} \cite{Bounouh1996} for the 
Co/Au/Co(0001) with a Au(111) overlayer. In both cases, the observed period(s) 
for the oscillations versus overlayer thickness were found to be in good 
agreement with the theoretically predicted ones. This effect has also been 
confirmed by means of first-principles calculations for the Co/Cu/Co(001) 
system with 
various types of overlayers \cite{Kudrnovsky1997, Kudrnovsky1997b, 
Kudrnovsky1998}. The comparison between the periods of 
oscillations versus overlayer thickness predicted theoretically and those 
observed experimentally is given in Table~\ref{ tab_comp}. A more detailed 
discussion of this effect can be found in Ref.~\cite{Bruno1996, 
Kudrnovsky1997, Kudrnovsky1998}.

\section{Strength and phase of interlayer exchange coupling}
\label{sec:strength}

In contrast with the excellent agreement between theory and experiment 
which is obtained for oscillation periods, the situation for the amplitude 
and phase of oscillations is less satisfactory. According to the theory 
exposed above, the coupling takes the following form in the limit of large 
spacer thickness (asymptotic limit):
\begin{equation}
J = \sum_\alpha \frac{A_\alpha}{D^2}\, 
\sin\left( q_\alpha D + \phi_\alpha \right) .
\end{equation}
Since the coupling constant $J$ has the dimension of an energy per unit 
area, the parameters $A_\alpha$ characterizing the coupling strength of the 
various oscillatory components have the dimension of an energy. By taking 
typical values of the Fermi wavevector and velocity, it is easy to see from 
eq.~(\ref{eq:J_asympt}) that they are typically of the order of 1 to 10~meV. 

Table~\ref{tab:amplitude} presents a comparison of theoretical and 
experimental values of the oscillation amplitude strengths $A_\alpha$, 
for various systems.
\footnote{Note that, in order to be able to compare 
various theoretical results with each other, we included in the present 
discussion only the calculations pertaining to semi-infinite magnetic 
layers.} 
We observe that there is a rather strong discrepancy 
between theory and experiment, but also among various theoretical studies. 
Although the agreement seems to be rather good for the Co/Cu(111)/Co 
system, more experimental and theoretical data would be required in order 
to know whether the apparent agreement is conclusive or accidental.

\subsection{Co/Cu(001)/Co}

The Co/Cu(001)/Co system is the one which 
has been most investigated theoretically and is considered to be a model 
system to test the predictions of theory. The theoretical results reported 
in Table \ref{tab:amplitude} correspond to semi-infinite magnetic layers, 
whereas the experimental data have been obtained for magnetic layers of 
finite thickness. As discussed in Section \ref{sec:maglayer} the strength 
of the coupling varies with magnetic layer thickness, which can be a source 
of discrepancy between theoretical an experimental results. An other possible 
source of discrepancy arise from unavoidable imperfections (roughness, 
intermixing) of the experimental samples.

Let us first address the short-period oscillatory component (labeled with  
the subscript 1). As discussed in Section \ref{sec:asympt} above, this 
component arise from 4 equivalent in-plane wavevectors ${\bf k}_{\|1}$ 
located on the 
$\overline{\Gamma}-\overline{X}$ high-symmetry line of the two-dimensional 
Brillouin zone \cite{Bruno1995}. Since the majority-spin band structure of 
fcc Co matches well the one of Cu, one has $|r_1^\uparrow| \approx 0$. 
On the other hand, for minority-spin fcc Co, there is a local gap in the band 
structure of symmetry compatible with the Cu states, which leads to 
total reflection, i.e., $|r_1^\downarrow| = 1$. Thus, one has 
$|\Delta r_1| \approx 0.5$ \cite{Lee1995, Stiles1999} and $|\Delta r_1|$ is 
(almost) 
independently of Co thickness \cite{Drchal1996}. The various theoretical 
values for the amplitude $A_1$ listed in Table \ref{tab:amplitude} 
agree rather well with each other, except 
the one of Ref. \cite{Stiles1996} which is almost a factor 2 larger than the 
values obtained by other authors \cite{Lee1995, Drchal1996, Mathon1997}. This 
discrepancy may be due to an error in the estimation of the curvature radius 
$\kappa_1$ of the Fermi surface, and of the Fermi velocity $v_{\bot 1}$, 
which are quite tricky to obtain accurately for ${\bf k}_{\| 1}$.

Turning now to the comparison between theory and experiment, we notice that 
the calculated values of $A_1$ are considerably larger than the measured 
ones. At least two reasons can contribute to this discrepancy. The first one 
is the effect of interface roughness, which generally tends to reduce the 
amplitude of the coupling oscillations \cite{Bruno1991}; this effect 
is particularly 
pronounced for short-period oscillatory components, as is indeed confirmed 
experimentally \cite{Weber1995}. The second reason is of 
intrinsic character: the theoretical values of $A_1$ given in 
Table \ref{tab:amplitude} correspond to the asymptotic limit, whereas the 
experimental data have been obtained for spacer thicknesses below 15~AL. 
As appears clearly from Fig.~6a of Ref. \cite{Drchal1996} and from 
Fig. 13 (bottom) of Ref. \cite{Mathon1997}, the asymptotic regime is attained  
only for thicknesses above 20 to 40 AL; below, the envelop of the 
oscillations deviates significantly from a $D^{-2}$ behavior, and the apparent 
amplitude in the range relevant to experiments is typically a factor 2 smaller 
than the asymptotic amplitude. This preasymptotic correction is attributed 
to a strong energy dependence of $r_1^\downarrow$ \cite{Mathon1997}. 

Let us now discuss the long-period oscillatory component. As appears from 
Table \ref{tab:amplitude}, the situation is quite confusing: not only the 
various theoretical results disagree with each other, but some of them 
\cite{Lee1995, Stiles1996, Mathon1997} {\em underestimate\/} the coupling 
strength as compared to the experiment \cite{Vries1996, Johnson1992, 
Johnson1993}, a fact which cannot be explained by the effect of roughness 
or interdiffusion.

The long-period oscillatory component 
arises from the center $\overline{\Gamma}$ of the two-dimensional Brillouin 
zone. Here again, for the same reason as above, one has 
$|r_2^\uparrow| \approx 0$. The minority-spin reflection coefficient, on the 
other hand, is considerably smaller than for the short-period oscillation, and 
one has $|r_2^\downarrow| \approx 0.15$ \cite{Bruno1995}, so that 
$|\Delta r_2| \approx 0.05$ \cite{Bruno1995, Lee1995}. This very small 
spin-dependent confinement explains the very small values of $A_2$ obtained 
by the authors which rely on the asymptotic expression (\ref{eq:J_asympt}) 
obtained from the stationary phase approximation \cite{Lee1995, Stiles1996, 
Mathon1997}. However, as seen from Fig. 2 of 
Ref. \cite{Bruno1995b} and Fig.~2 of Ref. \cite{Stiles1996}, 
$r_2^\downarrow$ increases very strongly with ${\bf k}_\|$ and full reflection 
is reached at a distance $0.1 \times \pi /a$ from $\overline{\Gamma}$; indeed, 
the low reflectivity arises only in a narrow window around $\overline{\Gamma}$.
As discussed in Ref. \cite{Bruno1998}, this gives rise to a strong 
preasymptotic correction, and explains why the stationary phase approximation 
approximation yields an {\em underestimated\/} value of $A_2$. On the other 
hand, if the ${\bf k}_\|$ integration is performed without using the stationary 
phase approximation, as was done in Ref. \cite{Drchal1996}, a much higher 
value of $A_2$ is obtained; the latter is larger than the experimental one 
\cite{Vries1996, Johnson1992, Johnson1993} by a factor 2.5, which seems 
plausible in view of the effect of roughness and interdiffusion.   

Our knowledge of the phase of the oscillations is much more restricted as 
this aspect of the problem has attracted little attention so far, with the 
notable exception of the work of Weber {\em et al.\/} \cite{Weber1995}. On 
general grounds, in the case of total reflection (as is the case for 
$r_{1\downarrow}$), one expects the phase to vary with magnetic 
layer thickness 
and/or with chemical nature of the magnetic layer; conversely, for a case 
weak confinement (as is the case for $r_2^\downarrow$), one expects the phase 
to be almost invariant \cite{Bruno1995}. These general trends are indeed 
confirmed experimentally by Weber {\em et al.\/} \cite{Weber1995}.

\begin{table}
\caption{Comparison between the theoretical predictions and 
experimental observations for the oscillation amplitudes $A_\alpha$ 
of interlayer exchange coupling versus spacer thickness; for Cu(001) 
and Au(001) spacers, $A_1$ and $A_2$ correspond, respectively, to the 
short-period and long-period oscillations.}
\label{tab:amplitude}
\vspace*{0.5\baselineskip}
\lineup
\begin{tabular}{@{}c|lc|lc@{}}
\br
system & theory & Ref. & experiment & Ref. 
\\
\mr
Co/Cu(111)/Co & \0$A_{\0} \approx$ \03.7\0 meV & \cite{Lee1995} & 
\0$A_{\0} \approx$ 7.6 meV 
& \cite{Johnson1992} \\
           & \0$A_{\0} \approx$ \04.2\0 meV & \cite{Stiles1996} & 
\0$A_{\0} \approx$ 3.4 meV 
& \cite{Schreyer1993} \\
           &                     &                   & 
\0$A_{\0} \approx$ 2.5 meV 
& \cite{Ives1994} \\
\mr
Co/Cu(001)/Co & 
$\begin{array}{l}
A_1 \approx 42 \mbox{\0\0\,\, meV} \\
A_2 \approx \00.13 \mbox{ meV}
\end{array}$ &
\cite{Lee1995} & 
$\begin{array}{l}
A_1 \approx 1.6 \mbox{ meV} \\
A_2 \approx 1.4 \mbox{ meV}
\end{array}$ & 
\cite{Vries1996, Johnson1992, Johnson1993} \\
\cline{2-2}
 & 
$\begin{array}{l}
A_1 \approx 72 \mbox{\0\0\,\, meV} \\
A_2 \approx \00.75 \mbox{ meV}
\end{array}$ &
\cite{Stiles1996} &
 &  \\
\cline{2-2}
 & 
$\begin{array}{l}
A_1 \approx 35 \mbox{\0\0\,\, meV} \\
A_2 \approx \03.5 \mbox{\0 meV}
\end{array}$ &
\cite{Drchal1996} & &  \\
\cline{2-2}
 & 
$\begin{array}{l}
A_1 \approx 35 \mbox{\0\0\,\, meV} \\
A_2 \approx \00.035 \mbox{ meV}
\end{array}$ &
\cite{Mathon1997} & &  \\
\mr
Fe/Au(001)/Fe & 
$\begin{array}{l}
A_1 \approx 12.5 \mbox{\0 meV} \\
A_2 \approx \06.9 \mbox{\0 meV}
\end{array}$ &
\cite{Stiles1996} &
$\begin{array}{l}
A_1 \approx 8.1 \mbox{ meV} \\
A_2 \approx 1.1 \mbox{ meV}
\end{array}$ & 
\cite{Unguris1997} \\
\br
\end{tabular}
\end{table}

\subsection{Fe/Au(001/Fe}

The system Fe/Au(001)/Fe is actually an excellent system for a quantitative 
test of the theory. This is due to the excellent lattice matching between 
Au and bcc Fe (with rotating the cubic axes of the latter 
by 45$^{\rm o}$), and to the 
availability of extremely smooth Fe substrates (whiskers) \cite{Unguris1994, 
Unguris1997}. 

In contrast to the Co/Cu(001) case discussed above, for Fe/Au(001), one 
has total reflection of minority-spin electrons both at ${\bf k}_{\| 1}$ 
(short-period oscillation) and ${\bf k}_{\| 2}$ (long-period oscillation), 
and $|r^\downarrow|$ is almost independent of ${\bf k}_\|$ around these 
points, as appears clearly from Fig.~1 of Ref. \cite{Stiles1999}. 
Therefore, the associated preasymptotic correction should not be 
very strong. 

Indeed, as is seen from Table \ref{tab:amplitude}, the predicted amplitudes 
are quite large, both for the short-period and long-period oscillatory 
components \cite{Stiles1996}. These predictions are fairly well confirmed 
by state-of-the-art experimental studies \cite{Unguris1997}, although the 
predicted amplitude of the long-period component is too large by a factor 6. 

Clearly, even for this almost ideal system, further work is required to 
achieve a satisfactory quantitative agreement between theory and experiment.

\section{Concluding remarks}

As has been discussed in detail in this review, there is a great deal of 
experimental evidence that the mechanism of quantum confinement presented 
above is actually the appropriate one to explain the phenomenon of 
oscillatory interlayer exchange coupling. This mechanism is entirely based 
upon a picture of independent electrons. This may seem paradoxal at first 
sight, in view of the fact that exchange interactions are ultimately due 
to the Coulomb interaction between electrons. In fact, this 
independent-electron picture can be be justified theoretically and is 
based upon the ``magnetic force theorem.'' A thorough discussion of this 
fundamental (but somehow technical) aspect of the problem is given 
elsewhere \cite{Bruno1999, Bruno1999b}.

In spite of the successes encountered by the quantum confinement mechanism, 
a number of questions remain to be clarify for a full 
understanding of the phenomenon. In particular, one needs to assess in a 
more quantitative manner than has been done so far the validity of the 
asymptotic expression (\ref{eq:J_asympt}); a first attempt towards 
addressing 
this issue is given in Ref. \cite{Bruno1998}.

\section*{Acknowledgements}

I am grateful to Claude Chappert, Josef Kudrnovsk\'y, Vaclav Drchal and 
Ilja Turek for their collaboration on the work presented in this paper.

\end{document}